\documentstyle[preprint,eqsecnum,aps]{revtex}
\tighten

\begin{document}
\draft

\preprint{SLAC-PUB-5998}

\title{
Relativistic Constituent Quark Model of Electroweak Properties of Baryons
}
\author{Felix Schlumpf}
\address{
Stanford Linear Accelerator Center\\
Stanford University, Stanford, California 94309
}
\date{\today}
\maketitle

\begin{abstract}
We calculate the electroweak properties of nucleons and hyperons in a
relativistic constituent quark model using the light-front formalism. The
parameters of the model, namely the constituent quark mass and the
confinement scale, can be uniquely chosen for both the electromagnetic and
weak experimental data. A consistent physical picture of the $qqq$ system
appears in this work with a symmetric nucleon wave function and an
asymmetric hyperon wave function. Only for the strangeness-changing weak
decays do we need nontrivial form factors of the constituent quark.
\end{abstract}

\pacs{11.10.St, 12.40.Aa, 13.30.Ce, 13.40.Fn}

\narrowtext

\section{Introduction}\label{sec:1}

The purpose of this paper is to present the results of comprehensive
calculations of electromagnetic and weak form factors of the baryon octet
in a relativistic constituent quark model.
This model was first formulated by Terent'ev and Berestetskii
\cite{bere76}, and has been applied to various hadronic processes by
Ref.
\cite{azna82,azna84}. Recently, new studies have been carried out
by Jaus in the meson sector \cite{jaus90} and by
Chung and Coester on the electromagnetic form factors of the nucleons
\cite{chun91}.
Nonrelativistic constituent quark models are successful in describing the
mass spectrum of baryons \cite{luch91}, but the use of nonrelativistic
quantum mechanics in calculating electroweak properties is inconsistent
when the mass of the quark is not large compared to the reciprocal
confinement scale.

In a relativistic theory the Poincar\'{e} invariance has to be
respected; this means, on the quantum level, the fulfillment of the
commutation relations between the generators of the Poincar\'{e} group.
Dirac \cite{dira49} has given a general formulation of methods to
satisfy simultaneously the requirements of special relativity and
Hamiltonian quantum mechanics. An extension of the Dirac classes of dynamics
can be found in Ref.~\cite{leut78}. The light-front scheme is in
particular distinguished from the other Dirac classes. Among the ten
generators of the Poincar\'{e} group, there are in the light-front
approach seven generators of kinematical character, and
only the remaining three generators contain
interaction, which is the minimal possible number.
The light-front dynamics is
therefore the most economical scheme for dealing with a relativistic
system. If we introduce the light-front variables $p^\pm\equiv p^0\pm p^3$,
the Einstein mass relation $p_\mu p^\mu = m^2$ is linear in $p^-$ and
linear in $p^+$, in contrast to the quadratic form in $p^0$ and $\vec p$ in
the usual dynamical scheme. A consequence is a single solution of the mass
shell relation in terms of $p^-$, in contrast to two solutions for $p^0$:
\[
p^- = (p_\perp^2 + m^2)/p^+\;, \qquad p^0 = \pm \sqrt{\vec p\,^2 + m^2}\;.
\]
The quadratic relation of $p^-$ and $p_\perp \equiv (p^1,p^2)$
in the above Equation
resembles the nonrelativistic scheme \cite{suss68}, and the variable $p^+$
plays the role of ``mass'' in this nonrelativistic analogy. It is therefore
a good idea to introduce relative variables like the Jacobi momenta
when dealing with several particles. As in the nonrelativistic scheme
such variables allow us to decouple the center of mass motion from the
internal dynamics. Hence we do not have the problems with the center of
mass motion which occur in the bag model. The light-front scheme shows another
attractive feature that it has in common with the infinite momentum
technique \cite{wein66}. In terms of the old fashioned (Heitler type, time
ordered, pre-Feynman) perturbation theory, the diagrams with quarks
created out of or annihilated into the vacuum do not contribute. The
usual $qqq$ quark structure is therefore conserved as in the
nonrelativistic theory. It is, however, harder to get the hadron states to
be eigenfunctions of the spin operator \cite{melo74}.

The equation of motion of the three-quark bound state on the light-front
can be reduced to a relativistic Schr\"odinger equation with an
effective potential. Since wave functions, that are solutions of the
relativistic Schr\"odinger equation, are not available, we start with two
simple baryon wave functions. The constituent quark mass $m_q$, the
length scale parameters $\beta$ and the quark form factors for
the weak decay are the parameters of this model. They are fixed by fitting
the relevant experimental data.

A consistent physical picture appears in this paper. The nucleon
consists of a symmetric three-quark state, whereas the
wave functions of the hyperons are asymmetric with a diquark
forming spin-0, V-spin-0 states and spin-0, U-spin-0 states,
respectively. Only for the strangeness-changing
weak decay do we need nontrivial form factors.
Recent works \cite{dzie} have also found evidence for diquark
clustering in the baryons.

In Sec.~\ref{sec:2} we give a brief summary of the light-front formalism
for the three-body bound state. Sec.~\ref{sec:3} contains
explicitly the asymmetric wave function on the light-front. We
discuss the different choices for the Ansatz of the wave function.
The magnetic moments of the baryon octet are calculated in
Sec.~\ref{sec:4} and the hyperon semileptonic weak decays are presented in
Sec.~\ref{sec:5}. We summarize our investigation in a concluding
Sec.~\ref{sec:6}. In the Appendix we give the connection between the
wave function and the effective potential.

\section{Light-front formalism for a three-body bound state}\label{sec:2}

To specify the dynamics of a many-particle system one
has to express the ten generators of the Poincar\'{e} group $P_\mu$ and
$M_{\mu\nu}$ in terms of dynamical variables. The kinematic subgroup
is the set of generators that are independent of the interaction. There
are five ways to choose these subgroups \cite{leut78}.
Usually a physical state is defined
at fixed $x_0$, and the corresponding hypersurface is left
invariant under the kinematic subgroup.

We shall use the light-front formalism which is specified by the invariant
hypersurface $x^+ = x^0+x^3 =$ constant. The following notation is used: The
four-vector is given by $x = (x^+,x^-,x_\perp)$, where $x^\pm = x^0
\pm x^3$ and $x_\perp=(x^1,x^2)$.
Light-front vectors are denoted by an arrow $\vec x =
(x^+,x_\perp)$, and they are covariant under kinematic Lorentz
transformations \cite{chun88}. The three momenta $\vec p_i$ of the quarks
can be transformed to the total and relative momenta to facilitate
the separation of the center of mass motion
\cite{bakk79}.

\begin{eqnarray}
\vec P&=&\vec p_1+\vec p_2+\vec p_3, \quad \xi={p_1^+\over p_1^++p_2^+}\;,
\quad
\eta={p_1^++p_2^+\over P^+}\;,\nonumber\\
&&\\
q_\perp&=&(1-\xi)p_{1\perp}-\xi p_{2\perp}\;, \quad
Q_\perp =(1-\eta)(p_{1\perp}+p_{2\perp})-\eta p_{3\perp}\;.\nonumber
\end{eqnarray}
Note that the four-vectors are not conserved, i.e., $p_1+p_2+p_3\not= P$.
In the light-front dynamics the Hamiltonian takes the form
\begin{equation}
H={P^2_\perp +\hat M^2 \over 2P^+}\;,
\end{equation}
where $\hat M$ is the mass operator with the interaction term $W$
\begin{eqnarray}
\hat M &=&M+W\;, \nonumber\\
M^2&=&{Q_\perp^2\over \eta(1-\eta)}+{M_3^2\over \eta}+{m_3^2\over 1-\eta},
\label{eq:2.3} \\
M_3^2&=&{q_\perp^2\over \xi (1-\xi)}+{m_1^2 \over \xi}+{m_2^2\over 1-\xi}\;,
\nonumber
\end{eqnarray}
with $m_i$ being the masses of the constituent quarks. To get a clearer
picture of $M$ we transform to $q_3$ and $Q_3$ by
\begin{eqnarray}
\xi&=&{E_1+q_3\over E_1+E_2}\;, \quad \eta={E_{12}+Q_3\over E_{12}+E_3}\;,
\nonumber\\
&&\\
E_{1/2}&=&({\bf q}^2+m_{1/2}^2)^{1/2}\;,\quad
E_{3}=({\bf Q}^2+m_{3}^2)^{1/2}\;,\quad
E_{12}=({\bf Q}^2+M_{3}^2)^{1/2}\;,\nonumber
\end{eqnarray}
where ${\bf q}=(q_1,q_2,q_3)$, and ${\bf Q}=(Q_1,Q_2,Q_3)$.
The expression for the mass operator is now simply
\begin{equation}
M=E_{12}+E_3\;, \quad M_3=E_1+E_2\;.
\end{equation}

The diagrammatic approach to light-front theory is well known
\cite{lepa80,mich82}. It provides in principal a framework
for a systematic treatment of higher-order gluon exchange. In
this work we limit ourselves to the tree graph. Since we set
$K^+=0$ we can preserve the correct $qqq$ structure of
the vertex. All relevant matrix elements we investigate are
related to
\begin{equation}
\left< \vec p\,'\left|\bar q \gamma^+ q\right| \vec p\right>
\sqrt{P^{'+}P^+} \equiv M^+ ,
\end{equation}
where the state $|\vec p\,\rangle\equiv|p\rangle/\sqrt{p^+}$ is
normalized according to
\begin{equation}
\left< \vec p\,' | \vec p \,\right>=\delta (\vec p\,' -\vec p\,) .
\end{equation}
By writing down the tree graph for the matrix element in light-front
variables for $K^+=0$, integrating over the ``--''
component of the loop-variables by
contour methods, and replacing vertex functions by wave functions (see
Appendix), we end up with the expression:
\begin{equation}
M^+=3{N_c\over (2\pi)^6}\int d^3qd^3Q\left({E'_3E'_{12}M\over E_3E_{12}M'}
\right)^{1/2}\Psi^\dagger({\bf q}',{\bf Q}',
\lambda')\Psi({\bf q},{\bf Q},\lambda)\;.
\label{eq:2.25}
\end{equation}

\section{Wave function models for the baryon octet}\label{sec:3}

In the light-front variables one can separate the center of mass motion
from the internal motion. The wave function $\Psi$ is therefore a
function of the relative momenta ${\bf q}$ and ${\bf Q}$. The product
$\Psi = \Phi\chi\phi$ with $\Phi =$ flavor, $\chi=$ spin, and $\phi=$
momentum distribution, is a symmetric function. We consider wave
functions $\Psi$ with spin-0, isospin-0 diquarks, with spin-0, V-spin-0
diquarks, and with spin-0, U-spin-0 diquarks, respectively. We write the
proton wave function as ($N_p$ being the normalization for the
proton):
\begin{equation} \label{eq:proton}
\left| p\right> =	N_p\left[
	-uud\left( \phi_1\chi^{\rho1}+\phi_2\chi^{\rho2}\right)
	+udu\left( \phi_1\chi^{\rho1}-\phi_3\chi^{\rho3}\right)
	+duu\left( \phi_2\chi^{\rho2}+\phi_3\chi^{\rho3}\right)
	\right] \; .
\end{equation}
The specific forms of the momentum wave functions $\phi_i$ and spin
$\chi$ are described below. The $\Lambda$ wave function is given by
\begin{equation} \label{eq:lambda}
\left| \Lambda\right> =	N_\Lambda\left[
	\phi_3\chi^{\rho3}\left( uds-dus \right)
	+\phi_2\chi^{\rho2}\left( usd-dsu \right)
	+\phi_1\chi^{\rho1}\left( sud-sdu \right)
	\right] \; .
\end{equation}
The wave functions for the other members of the baryon octet are obtained
by changing the flavor wave function appropriately, for instance:
\begin{eqnarray}
| n \rangle &=& - | p \rangle \left( u \leftrightarrow d \right)\;,\nonumber\\
| \Sigma^+ \rangle &=& - | p \rangle \left( d \rightarrow s \right)\;.
\end{eqnarray}

The angular momentum ${\bf j}$ can be expressed as a sum of orbital and
spin contributions
\begin{eqnarray}
{\bf j}=i\nabla_{\bf p}\times {\bf p}+\sum_{j=1}^3 {\cal R}_{Mj}{\bf s}_j \;,
\end{eqnarray}
where ${\cal R}_M$ is a Melosh rotation
acting on the quark spins
${\bf s}_j$, which has the matrix representation (for two particles)
\begin{equation}
\left< \lambda' |{\cal R}_M(\xi,q_\perp,m,M)|\lambda\right> =
\left[ {m+\xi M-i\bbox{\sigma}\cdot({\bf n}\times {\bf q})\over
\sqrt{(m+\xi M)^2+q_\perp^2}}\right]_{\lambda'\lambda}
\end{equation}
with ${\bf n}=(0,0,1)$. In previous works
\cite{dzie88} this
rotation has been approximated by putting $M=M_B$. This corresponds to a
weak-binding limit which cannot be justified for a bound state in QCD.
In this limit our model has a close connection to many other
relativistic quark models as shown by Koerner et al. \cite{korn91}.

The operator ${\bf j}$ commutes with the mass operator $\hat M$; this is
necessary and sufficient for Poincar\'e-invariance of the bound state.

In terms of the relative momenta the angular momentum takes the
form
\begin{eqnarray}
{\bf j}&=& i\nabla_{\bf Q}\times {\bf Q}+{\cal R}_M(\eta,Q_\perp,M_3,M)
{\bf j}_{12}+{\cal R}_M(1-\eta,-Q_\perp,m_3,M){\bf s}_3\;,\nonumber\\
&&\\
{\bf j}_{12}&=& i\nabla_{\bf q}\times {\bf q}+{\cal R}_M(\xi,q_\perp,m_1,M_3)
{\bf s}_1+{\cal R}_M(1-\xi,-q_\perp,m_2,M_3){\bf s}_2\;.\nonumber
\end{eqnarray}
We can drop the orbital contribution to obtain:
\begin{eqnarray}
{\bf j}&=&\sum {\cal R}_i{\bf s}_i\;,\nonumber\\
{\cal R}_1&=&{1\over \sqrt{a^2+Q_\perp^2}\sqrt{c^2+q_\perp^2}}
\pmatrix{ac-q_RQ_L&-aq_L-cQ_L\cr
         cQ_R+aq_R&ac-q_LQ_R}\;,\nonumber\\
{\cal R}_2&=&{1\over \sqrt{a^2+Q_\perp^2}\sqrt{d^2+q_\perp^2}}
\pmatrix{ad+q_RQ_L&aq_L-dQ_L\cr
         dQ_R-aq_R&ad+q_LQ_R}\;,\label{eq:melosh}\\
{\cal R}_3&=&{1\over \sqrt{b^2+Q_\perp^2}}\pmatrix{b&Q_L\cr
         -Q_R&b}\;,\nonumber
\end{eqnarray}
with
\begin{eqnarray}
a&=&M_3+\eta M\;,\quad b=m_3+(1-\eta)M\;,\nonumber\\
c&=&m_1+\xi M_3\;, \quad d=m_2+(1-\xi)M_3\;,\nonumber\\
q_R&=&q_1+iq_2\;,\quad q_L=q_1-iq_2\;,\\
Q_R&=&Q_1+iQ_2\;,\quad Q_L=Q_1-iQ_2\;.\nonumber
\end{eqnarray}

The spin wave functions in Eqs.~(\ref{eq:proton}) and (\ref{eq:lambda}) are:
\begin{eqnarray}
\label{eq:spinfunction}
\chi^{\rho3}_\uparrow&=&{1\over\sqrt 2}(\uparrow\downarrow\uparrow-
\downarrow\uparrow\uparrow)\;,\nonumber\\
\chi^{\rho3}_\downarrow&=&{1\over\sqrt 2}(\uparrow\downarrow\downarrow
-\downarrow\uparrow\downarrow)\;.
\end{eqnarray}
with $\chi^{\rho2}$ and
$\chi^{\rho1}$ being the appropriate permutations of $\chi^{\rho3}$.
The spin-wave function of the $i$th quark is given by
\begin{equation}
\uparrow={\cal R}_i\pmatrix{1\cr 0} \hbox{  and  }
\downarrow={\cal R}_i\pmatrix{0\cr 1}\;.
\end{equation}

The functions $\phi_i$ in Eqs.~(\ref{eq:proton}) and (\ref{eq:lambda})
are the momentum wave functions symmetric in the quarks different form
the $i$th quark. We choose a harmonic oscillator and a pole type wave
function:
\begin{eqnarray}
\phi_i^H &=& e^{-X_i} ,\label{eq:harmon}\\
\phi_i^P &=& (1+X_i)^{-3.5} \label{eq:pole},
\end{eqnarray}
where the $X_i$ are the generalized forms of $M^2/2\beta^2$:
\begin{eqnarray} \label{eq:difunction}
X_3 & = &
\frac{Q_\perp^2}{2\eta(1-\eta)\beta_Q^2}+\frac{q_\perp^2}
{2\eta\xi(1-\xi)\beta_q^2}+\frac{m_1^2}{2\eta\xi\beta_q^2}+\frac{m_2^2}
{2\eta(1-\xi)\beta_q^2}+\frac{m_3^2}{2(1-\eta)\beta_Q^2}
\; ,
\nonumber \\
 X_2 & = &
q_\perp^2\frac{(1-\eta)(1-\xi)\beta_Q^2+\xi\beta_q^2}{2\beta_Q^2\beta_q^2
\eta\xi(1-\xi)(1-\eta+\xi\eta)}+
Q_\perp^2\frac{(1-\xi)(1-\eta)\beta_q^2+\xi\beta_Q^2}{2\beta_Q^2\beta_q^2\eta
(1-\eta)(1-\eta+\xi\eta)}
\nonumber \\
 &  &+
q_\perp Q_\perp \frac{\beta_Q^2-\beta_q^2}{\beta_Q^2
\beta_q^2\eta(1-\eta+\xi\eta)}+
\frac{m_1^2}{2\eta\xi\beta_q^2}+\frac{m_2^2}{2\eta(1-\xi)\beta_Q^2}+
\frac{m_3^2}{2(1-\eta)\beta_q^2} \; ,
\nonumber \\
 X_1 & = &
q_\perp^2\frac{(1-\xi)\beta_q^2+\xi(1-\eta)\beta_Q^2}{2\beta_Q^2\beta_q^2
\eta\xi(1-\xi)(1-\xi\eta)}+Q_\perp^2\frac{(1-\xi)\beta_Q^2+\xi(1-\eta)
\beta_q^2}{2\beta_Q^2\beta_q^2\eta(1-\eta)(1-\xi\eta)}
\nonumber \\
 &  &-
q_\perp Q_\perp \frac{\beta_Q^2-\beta_q^2}
{\beta_Q^2\beta_q^2\eta(1-\xi\eta)}+
\frac{m_1^2}{2\eta\xi\beta_Q^2}+\frac{m_2^2}{2\eta(1-\xi)\beta_q^2}+
\frac{m_3^2}{2(1-\eta)\beta_q^2} \; .
\end{eqnarray}

The normalization factors $N_B$ in
Eqs.~(\ref{eq:proton}) and (\ref{eq:lambda}) are determined from $\langle
B|B\rangle=1$. For the proton and $\Lambda$ do we get:
\begin{eqnarray}
\langle p|p\rangle &=& N_p^2 \left[
(\phi_1^2+\phi_2^2+\phi_1\phi_2)_{uud}+
(\phi_1^2+\phi_3^2+\phi_1\phi_3)_{udu}+
(\phi_2^2+\phi_3^2+\phi_2\phi_3)_{duu} \right] ,\nonumber\\
\langle \Lambda|\Lambda\rangle &=& N_\Lambda^2
\left[(\phi_1^2)_{sud}+
(\phi_2^2)_{usd}+(\phi_3^2)_{uds}\right] .\nonumber
\end{eqnarray}

Our wave functions $\phi^H$ and $\phi^P$ only differ in their high energy
behavior. The exponent of the pole type wave function can be chosen by
fitting the electromagnetic form factors of the nucleons \cite{schl92b}.
In the limit of vanishing quark masses do the corresponding quark
distribution amplitudes both converge to the asymptotic form of
Ref.~\cite{lepa80}:
\begin{equation}
\phi^H, \phi^P \propto \xi\eta^2(1-\eta)(1-\xi) = x_1 x_2 x_3 ,
\end{equation}
with the light-front fractions $x_i\equiv p_i^+/P^+$.

\section{Magnetic Moments}\label{sec:4}

The electromagnetic current matrix element for the transition
$B \to B'\gamma$ can be written
in terms of two form factors taking into account current and parity
conservation:

\begin{equation}
\left< B',\lambda ' p' \left| J^\mu \right|
B,\lambda p\right> =
\bar u_{\lambda '}(p') \left[ F_1(K^2)\gamma^\mu + {F_2(K^2) \over
2 M_N}i\sigma^{\mu\nu}K_\nu \right] u_\lambda (p)
\label{eq:3.1}
\end{equation}
with momentum transfer $K = p' - p$, and the current $J^\mu=
\bar q e \gamma^\mu q$.
In order to use Eq. (\ref{eq:2.25}) we express the form factors in terms of the
$+$ component of the current:
\begin{eqnarray}
F_1(K^2) &=&\left< B',\uparrow\left| J^+\right| B,
\uparrow\right>\;,\nonumber\\
&&\\
K_\perp F_2(K^2) &=&-{2M_N}\left< B',\uparrow\left|
J^+\right| B,\downarrow\right>\;.\nonumber
\end{eqnarray}
For $K^2 = 0$ the form factors $F_1$ and $F_2$ are respectively equal
to the charge and the anomalous magnetic moment in units $e$ and
$e/M_N$, and the magnetic moment is $\mu = F_1(0) + F_2(0)$.

The anomalous magnetic moment for the $\Lambda$ is given by:
\begin{equation} 
K_\perp F_2(0)=-2M_\Lambda N_\Lambda^2{N_c\over (2\pi)^6} \int d^3q d^3Q
\sum_{i=1}^3 e_i |\phi_i|^2 \left< \chi_\uparrow^{\rho i}
| \chi_\downarrow^{\rho i} \right> \; ,
\end{equation}
with $e_i$ being the charge of the $i$th quark. The formulae for the other
members of the baryon octet are analogous. The calculation of the spin
matrix elements is tedious but straightforward. The explicit expressions
are given in Ref.~\cite{schl92a}. The numerical results are summarized in
Table~\ref{table:2} for the four different parameter sets given in
Table~\ref{table:1}. Parameter sets~1 and 2 are given as a reference for
symmetric wave functions ($\beta_q=\beta_Q$), which are usually used in the
literature \cite{chun91,tupp88}. Set~1 uses relatively large quark masses
normally found in nonrelativistic models ($m_u=m_d\approx
M_{\text{Nucleon}}/3$).
The magnetic moments can be reproduced very well, but the semileptonic
weak decay data deviate by more than an order of magnitude. This is due
to the special choice of the $\beta$ parameter for the nucleon and the
hyperons:
\begin{equation}
\beta_N \ll \beta_\Sigma \approx \beta_\Lambda \approx \beta_\Xi\;,
\end{equation}
which results in a too large suppression for the $\Delta$S=1 transitions,
since the wave function overlap is small. Parameter set~2 on the other hand
gives good values for the semileptonic decays, but is bad at fitting the
magnetic moments. Within the symmetric wave function model do we find,
that either the magnetic moments can be fitted and the weak decay data are
poorly fitted or vice versa. The opposite statement in Ref.~\cite{azna84}
has to be questioned, because their numerical results for the magnetic
moments are wrong. Our results agree with Ref.~\cite{tupp88} on this point.
The inconsistency just described between the electromagnetic and the weak
sector can be resolved by using asymmetric wave functions (Parameter sets~3
and 4).
All electroweak properties in Table~\ref{table:2} can be fitted with this
wave function. Set~3 use the harmonic oscillator wave function in
Eq.~(\ref{eq:harmon}) and set~4 uses the pole type wave function in
Eq.~(\ref{eq:pole}). The only essential difference between these two
types of wave functions is the high energy behavior \cite{schl92b}. The
twelve parameters in Table~\ref{table:1} are overcounted because
$\beta_{qN}=\beta_{q\Lambda}=\beta_{ud}$ and
$\beta_{q\Sigma}=\beta_{q\Xi}=\beta_{us}$ being the scale parameters for
the diquarks $ud$ and $us$, respectively. After fitting the mass
$m_u=m_d$, we fix the strange quark mass to be $m_s/m_u\sim
1.4-1.6$
\cite{scad81}. Therefore, we have only nine degrees of freedom, but it is
not obvious that a reasonable fit is possible since the relations are
nonlinear. We get however an excellent agreement with data for the
asymmetric wave functions (sets~3 and 4). The neutron magnetic moment could
be improved by introducing electromagnetic quark form factors
\cite{chun91}.

\section{Hyperon Semileptonic Beta Decay}\label{sec:5}

In the low energy limit the standard model for semileptonic weak decays
reduces to an effective current-current interaction Hamiltonian
\begin{equation}
H_{\rm int} = {G \over \sqrt{2}} J_\mu L^\mu + {\rm h.c. }\;,
\end{equation}
where $G \simeq 10^{-5}/M_p^2$ is the weak coupling constant,
\begin{equation}
L^\mu = \bar{\psi_e} \gamma^\mu (1 - \gamma_5) \psi_\nu +
   \bar{\psi_\mu} \gamma^\mu (1 - \gamma_5) \psi_\nu
\end{equation}
is the lepton current, and
\begin{equation}
J_\mu = V_\mu - A_\mu\;, \quad
V_\mu = V_{ud} \bar u \gamma_\mu d + V_{us} \bar u \gamma_\mu s\;, \quad
A_\mu = V_{ud} \bar u \gamma_\mu \gamma_5 d + V_{us} \bar u \gamma_\mu
\gamma_5 s \;,
\end{equation}
is the hadronic current, and $V_{ud}, V_{us}$ are the elements of the
Kobayashi-Maskawa mixing matrix.
The $\tau$-lepton current cannot
contribute since $m_\tau$ is much too large.

The matrix elements of the hadronic current between spin-${1 \over 2}$
states are
\begin{equation}
\left< B',p' \left| V^\mu\right| B,p \right> = V_{qq'} \bar u(p') \left[
  f_1(K^2) \gamma^\mu - {f_2(K^2) \over M_i} i \sigma^{\mu\nu} K_\nu
  + {f_3(K^2) \over M_i}  K^\mu \right] u(p)\;,
\end{equation}
\begin{equation}
\left< B',p' \left| A^\mu\right| B,p \right> = V_{qq'} \bar u(p') \left[
  g_1(K^2) \gamma^\mu - {g_2(K^2) \over M_i} i \sigma^{\mu\nu} K_\nu
  + {g_3(K^2) \over M_i}  K^\mu \right] \gamma_5 u(p)\;,
\end{equation}
where $K = p - p'$ and $M_i$ is the mass of the initial baryon.
The quantities $f_1$ and $g_1$ are the vector and axial-vector
form factors, $f_2$ and $g_2$ are the weak magnetism and electric form
factors and $f_3$ and $g_3$ are the induced scalar and pseudoscalar form
factors, respectively. T invariance implies real form factors.
We do not calculate $f_3$ and $g_3$ since we put $K^+ = 0$
and their dependence on the decay spectra is of the order
\begin{equation}
\left( m_l \over M_i \right)^2 \ll 1\;,
\end{equation}
where $m_l$ is the mass of the final charged lepton.
The other form factors are
\begin{eqnarray}
f_1 &=&\left< B',\uparrow\left| V^+\right| B,\uparrow \right>\;,\nonumber\\
K_\perp f_2 &=&M_i\left< B',\uparrow\left| V^+\right|
B,\downarrow \right>\;,\nonumber\\
g_1 &=&\left< B',\uparrow \left| A^+\right| B,\uparrow\right>\;,\nonumber\\
K_\perp g_2 &=&-M_i\left< B',\uparrow \left| A^+\right| B,\downarrow\right>\;.
\label{eq:formfactors}
\end{eqnarray}

We generalize the Dirac quark current for the $s\to u$ transition
by introducing constituent quark form factors $f_{1us}$ and $g_{1us}$:
\[
\bar u\gamma^\mu(1-\gamma_5)s \rightarrow
\bar u\gamma^\mu(f_{1us} - g_{1us}\gamma_5)s .
\]
We therefore have an effective $\tilde f_1=f_{1us}f_1$ and an effective
$\tilde g_1=g_{1us}g_1$.
Ignoring the lepton-mass the rate $\Gamma$ is given by \cite{garc85}:
\begin{eqnarray}
\Gamma &=&
G^2 {\Delta M^5 |V|^2 \over 60\pi^3}\Bigl[ (1 - {3 \over 2} \beta +
  {6 \over 7} \beta^2) \tilde f_1^2 + {4 \over 7} \beta^2 f_2^2 + (3 -
  {9 \over 2} \beta + {12 \over 7} \beta^2) \tilde g_1^2  \nonumber\\
  &&+ {12 \over 7} \beta^2 g_2^2
  + {6 \over 7} \beta^2 \tilde f_1 f_2
  + (-4\beta + 6\beta^2) \tilde g_1 g_2
  + {4 \over 7} \beta^2 (\tilde f_1 \lambda_f
  + 5\tilde g_1\lambda_g) \Bigr]\;,
\label{eq:rate}
\end{eqnarray}
where $\beta$ is defined as $\beta = (M_i - M_f) / M_i$, and $\Delta M
=M_i-M_f$,
$M_i$, $M_f$ being the masses of the initial and final baryon,
respectively. The $K^2$-dependence of $f_2$ and $g_2$ is ignored and
$f_1$ and $g_1$ are expanded as
\begin{equation}
\tilde f_1(K^2) = \tilde f_1(0) + {K^2 \over M_i^2} \lambda_f\;, \quad
\tilde g_1(K^2) = \tilde g_1(0) + {K^2 \over M_i^2} \lambda_g \;.
\end{equation}
We correct the rate from Eq.~(\ref{eq:rate}) to include the effect
of the nonvanishing lepton mass and the effect of the radiative
corrections \cite{garc85,garc82}.

The form factors in Eq.~(\ref{eq:rate}) are calculated by using
Eqs.~(\ref{eq:formfactors}), (\ref{eq:proton}), and (\ref{eq:lambda}). As
an example we give the form factors $g_1$ for the decay $\Lambda \to p
l^-\bar \nu_l$:
\begin{eqnarray} 
	g_1	& = & \langle p,\uparrow | A^+|\Lambda,\uparrow \rangle=
	\nonumber \\
	&   & N_p N_\Lambda \left(
\phi_1\phi_3 \langle\chi_\uparrow^{\rho 3}|A^+|\chi_\uparrow^{\rho 1}\rangle
-\phi_2\phi_3 \langle\chi_\uparrow^{\rho 3}|A^+|\chi_\uparrow^{\rho 2}\rangle
-2\phi_3^2 \langle\chi_\uparrow^{\rho 3}|A^+|\chi_\uparrow^{\rho 3}\rangle
\right)
	\; .
\end{eqnarray}
The calculations for the spin matrix elements are tedious but simple
algebra. The exact formulae are given in Ref.~\cite{schl92a}. In the limit
of symmetric wave functions and $K^2=0$ do we get for the above decay
$\Lambda \to p l^-\bar \nu_l$:
\begin{equation} 
	g_1(0)=-\sqrt{\frac{3}{2}}\frac{N_c}{(2\pi)^6}\int d^3q d^3Q
	\frac{\phi^\dagger(M)\phi (M)
        (b'b-Q_\perp^2)(a'a+Q_\perp^2)^2}{(a'^2+Q_\perp^2)
	(a^2+Q_\perp^2)\sqrt{(b'^2+Q_\perp^2)(b^2+Q_\perp^2)}} \; .
\end{equation}
The $K^2$-dependence of the form factors $f_1$ and $g_1$ is calculated
by their derivatives at $K^2=0$. The form factor $g_2$ vanishes or is
very small as it should be. The weak magnetism form factor $f_2$ agrees
with the conserved vector current (CVC) hypothesis within 5\%. The form
factors $f_1$ and $g_1$ are given in Table~\ref{table:3} for the various
decays. We summarize the ratios $g_1/f_1$ and the rates $\Gamma$ for all
the measured semileptonic weak decays in Table~\ref{table:2}.
Sets~3 and 4 give an excellent fit
for all experimental data, except for the ratio $g_1/f_1$ for the decays
$\Lambda\to p$ and $\Sigma^-\to n$. This is however a general property of
every quark model due to its SU(6) flavor-spin symmetry. The ratio
\begin{equation} 
	\frac{g_1/f_1(\Lambda\to pe^-\bar \nu_e)}
	{g_1/f_1(\Sigma^-\to ne^-\bar \nu_e)}
\end{equation}
is constrained to be $-3$ in the models in contrast to the experimental
value $-2.11\pm 0.15$ for $g_2=0$.

\section{Summary and outlook}\label{sec:6}

Nonrelativistic constituent quark models for the electroweak properties
of the baryons are inconsistent even for small values of the
momentum transfer. We have shown that there exists a relativistic quark
model with diquark clustering that provides a framework,
in which we have overall an excellent and consistent picture of the
whole baryon octet for the magnetic moments and the semileptonic
weak decays.
The physical picture of the baryon octet is as follows (Parameter
sets~3 and 4 in Table~\ref{table:1}). There is no diquark
clustering in the nucleon sector ($\beta_{qN}=\beta_{QN}$). In
the strange sector we have a strong diquark clustering for the
$\Sigma$s ($\beta_{q\Sigma}\sim 2\beta_{Q\Sigma}$) and $\Xi$s
($\beta_{q\Xi}\sim 2\beta_{Q\Xi}$) and a small one for the $\Lambda$
($1.5 \beta_{q\Lambda}\sim \beta_{Q\Lambda}$). The diquark
$us$-pair ($\beta_{q\Sigma}, \beta_{q\Xi}$) is more tightly bound than
the $ud$-pair ($\beta_{qN}, \beta_{q\Lambda}$) as we might expect.
The low momentum properties do not depend on the two different
wave functions chosen. It would however be illuminating
to derive the momentum wave function from a potential. To complete
the study of the baryons one should also include the effects
of higher Fock states.

\acknowledgments

It is a pleasure to thank W.~Jaus for helpful discussions.
This work was supported in part by the Schweizerischer Nationalfonds and
in part by the Department of Energy, contract DE-AC03-76SF00515.

\appendix

\section*{Connection between the wave function and
the potential}\label{app:1}

It is instructive to give some details on the derivation of the equation
of motion for the wave function.

We shall assume only two-particle forces interacting in a ladder-type
pattern so that the dynamics of the three-body system is governed
by the Bethe-Salpeter (BS) interaction kernel
for the two-body system and
the relativistic Faddeev equations.

Using the Faddeev decomposition for the vertex function $\Gamma =
\Gamma^{(1)} +\Gamma^{(2)} +\Gamma^{(3)}$, we can write down a BS equation for
the various components in operator notation
\begin{equation}
\Gamma^{(1)}=T^{(1)}G_2G_3(\Gamma^{(2)}+\Gamma^{(3)})
\end{equation}
with
\begin{equation}
G_i=\not\! p_i - m_i, \quad T^{(1)}=(1-VG_2G_3)^{-1}V\;,
\end{equation}
and similarly for $\Gamma^{(2)}$ and $\Gamma^{(3)}$. $V$ is the one
gluon exchange kernel between two quarks, and $T$ is already the ladder
sum to all orders. It is useful to consider the second iteration of the
vertex equation, which is given by:
\begin{equation}
{\bf \Gamma} = U G_1 G_2 G_3 {\bf \Gamma}\;,
\label{eq:2.8}
\end{equation}
where ${\bf \Gamma}=(\Gamma^{(1)},\Gamma^{(2)},\Gamma^{(3)})$
and $U$ is the matrix
\begin{equation}
U_{ij}=\cases{T^{(i)}G_jT^{(k)}&for $i\neq j$ with $k\neq i,j$ ,\cr
              T^{(i)}(G_kT^{(l)}+G_lT^{(k)})&for $i=j$ with $k\neq l\neq
			  i$ .}
\end{equation}
The four-dimensional Eq.~(\ref{eq:2.8}) can be reduced to a three-dimensional
equation
\begin{equation}
{\bf \Gamma}=W g_3 {\bf \Gamma}\;, \quad W=(1-U R_3)^{-1}U
\end{equation}
by writing $G_1 G_2 G_3=g_3+R_3$ where $g_3$ has only three-particle
singularities.  We choose a $g_3$ which puts the quarks on
their mass shells:
\begin{equation}
g_3=(2\pi i)^2\int ds{1 \over P^2-s}\prod_{i=1}^3\delta^+(p_i^2-m_i^2)
(\not\! p_i + m_i)\;,
\end{equation}
where $P$ is the total momentum of the bound state, $s=(p_1+p_2+p_3)^2$
and $p_i$ are restricted by $p_i^+ \ge 0$. We get
\begin{equation}
g_3=(2\pi i)^2 \delta(p_2^2-m_2^2)\delta(p_3^2-m_3^2)\Theta(\xi)\Theta
(1-\xi)\Theta(\eta)\Theta(1-\eta){\Lambda^+(p_1)\Lambda^+(p_2)
\Lambda^+(p_3)\over \xi\eta (P^2-M^2)}
\label{eq:green}
\end{equation}
with the spin projection operator
\begin{equation}
\Lambda^+(p_i)=\sum_\lambda u(p_i,\lambda)\bar u(p_i,\lambda)\;.
\end{equation}
Writing
\begin{eqnarray}
\hat g_3&=&{1\over P^2-M^2}\;,\nonumber\\
\hat \Gamma^{(i)}&=&\left( {M_3M\over E_1E_2E_3E_{12}}\right)^{1/2}
\Gamma^{(i)}u(p_1\lambda_1)u(p_2\lambda_2)u(p_3\lambda_3)\;,\\
\hat W_{ij}&=&\left( {M_3M'_3MM'\over E_1E'_1E_2E'_2E_3E'_3E_{12}E'_{12}}
\right)^{1/2}u(p_1\lambda_1)u(p_2\lambda_2)u(p_3\lambda_3)W_{ij}
\bar u(p'_1\lambda'_1)\bar u(p'_2\lambda'_2)\bar u(p'_3\lambda'_3)
\nonumber
\end{eqnarray}
we are led to the integral equation
\begin{eqnarray}
\hat \Gamma^{(i)}({\bf q},{\bf Q},\lambda_1,\lambda_2,\lambda_3)&=
&{1\over (2\pi)^6}\sum_{\lambda'_1\lambda'_2\lambda'_3j}\int d^3\!q'd^3\!Q'
\hat W^{ij}({\bf q},{\bf q}',{\bf Q},{\bf Q}',\lambda_1,\lambda'_1,
\lambda_2,\lambda'_2,\lambda_3,\lambda'_3)\nonumber\\
&&\times \hat g_3({\bf q}',{\bf Q}') \hat \Gamma^{(j)}
({\bf q}',{\bf Q}',\lambda'_1,\lambda'_2,\lambda'_3)\;.
\end{eqnarray}
We can write this equation in terms of the wave function $\Psi$. The
Faddeev decomposition
is $\Psi=\Psi^{(1)}+\Psi^{(2)}+\Psi^{(3)}$,
the relation to the vertex function is $\Psi^{(i)}=\hat g_3
\hat \Gamma^{(i)}$, and writing ${\bf\Psi}=(\Psi^{(1)}, \Psi^{(2)},
\Psi^{(3)})$ we get
\begin{equation}
(M_B^2-M^2){\bf\Psi}=\hat W {\bf\Psi}
\end{equation}
with $M_B$ being the mass of the baryon. If we put $\hat W=MW+WM+W^2$
we see that the wave function is an eigenfunction of the mass operator
$\hat M^2$, given in Eq.~(\ref{eq:2.3}):
\begin{equation}
\hat M^2 \Psi = M^2_B \Psi
\end{equation}
which is equivalent to the equation usually used in constituent quark
models \cite{godf89}
\begin{equation}
(E_{12}+E_3+W)\Psi = M_B \Psi\;.
\label{eq:2.18}
\end{equation}
This last equation is the starting point for an explicit calculation of the
wave function, which has been done for the meson sector
\cite{jaco90}.

\narrowtext
\begin{table}
\caption{The parameters of the constituent quark model: quark masses $m$
(GeV), scale parameter $\beta$ (GeV) and quark form factors $f_1$ and
$g_1$ for the quark
transition $s\to u$. Note that sets~1--3 are used for the harmonic
oscillator wave function, whereas set~4 is used for the pole type
wave function.}
\begin{tabular}{ldddd}
&\multicolumn{3}{c}{Harmonic Oscillator}&\multicolumn{1}{c}{Pole Type}\\
Parameters& Set 1& Set 2 &Set 3 &Set 4  \\
\tableline
$m_u=m_d$ & 0.33 & 0.267 & 0.26 & 0.263 \\
$m_s$     & 0.55 & 0.40  & 0.38 & 0.38  \\
$\beta_{qN}$& 0.16 & 0.56 & 0.55 & 0.607 \\
$\beta_{QN}$& 0.16 & 0.56 & 0.55 & 0.607 \\
$\beta_{q\Lambda}$& 1.00 & 0.60 & 0.55 & 0.607 \\
$\beta_{Q\Lambda}$& 1.00 & 0.60 & 0.80 & 0.90 \\
$\beta_{q\Sigma}$& 1.00 & 0.60 & 0.80 & 0.90 \\
$\beta_{Q\Sigma}$& 1.00 & 0.60 & 0.40 & 0.45 \\
$\beta_{q\Xi}$& 1.08 & 0.62 & 0.80 & 0.90 \\
$\beta_{Q\Xi}$& 1.08 & 0.62 & 0.36 & 0.40 \\
$f_{1us}$     & 1.00 & 1.00 & 1.19 & 1.28 \\
$g_{1us}$     & 1.00 & 1.00 & 1.19 & 1.28 \\
\end{tabular}
\label{table:1}
\end{table}

\mediumtext
\begin{table}
\caption{Electroweak properties of the baryon octet. The calculations
with symmetric wave functions (sets~1 and 2) and asymmetric wave functions
(sets~3 and 4) are compared. Note that set~1 is only able to fit
the magnetic moments, whereas set~2 is best at fitting the weak decays.
Sets~3 and 4 reproduce all electroweak data in an excellent way. The
magnetic moments are given in units of the nuclear magneton; the decay
rates are given in units of $10^6s^{-1}$ (except the nucleon decay is in
units of $10^{-3}s^{-1}$). Experimental data are from Ref.~\protect
\cite{pdg92}.}
\begin{tabular}{ld@{${}\pm{}$}ldddd}
Quantity&\multicolumn{2}{c}{Expt.}& Set 1 & Set 2 & Set 3 & Set 4  \\
\tableline
$\mu(p)$ & 2.79 & $10^{-7}$ & 2.85 & 2.78 & 2.82 & 2.81 \\
$\mu(n)$ &--1.91 & $10^{-6}$ &--1.83 &--1.62 &--1.66 &--1.66 \\
$\mu(\Sigma^+)$ & 2.42 & 0.05 & 2.59 & 3.23 & 2.63 & 2.61 \\
$\mu(\Sigma^-)$ &--1.160 & 0.025 &--1.30 &--1.36 &--1.14 &--1.13 \\
$\mu(\Lambda)$ &--0.613 & 0.004 &--0.48 &--0.72 &--0.69 &--0.69 \\
$\mu(\Xi^0)$ &--1.250 & 0.014 &--1.25 &--1.87 &--1.25 &--1.24 \\
$\mu(\Xi^-)$ &--0.6507 & 0.0025 &--0.99 &--0.96 &--0.67 &--0.76 \\
$g_1/f_1(n\to p e^-\bar\nu_e)$ &1.2573&0.0028&1.63&1.252&1.248&1.260\\
$\sqrt{3/2}g_1(\Sigma^\pm\to \Lambda e^\pm\nu_e)$ &0.742&0.018&0.80
&0.736&0.759&0.704\\
$g_1/f_1(\Lambda\to p e^-\bar\nu_e)$ &0.718&0.015&0.957&0.826&0.759&0.745\\
$g_1/f_1(\Sigma^-\to n e^-\bar\nu_e)$ &--0.340&0.017&--0.319&--0.275&
--0.255&--0.255\\
$g_1/f_1(\Xi^-\to \Sigma^0e^-\bar\nu_e)$ &1.287&0.158&1.594&1.362&
1.212&1.192\\
$g_1/f_1(\Xi^-\to \Lambda e^-\bar\nu_e)$ &0.25&0.05&0.319&0.272&0.270&0.255\\
$\Gamma(n\to p e^-\bar\nu_e)$ &1.125&0.003&1.76&1.152&1.113&1.13\\
$\Gamma(\Sigma^+\to \Lambda e^+\nu_e)$ &0.25&0.06&0.29&0.24&0.25&0.21\\
$\Gamma(\Sigma^-\to \Lambda e^-\bar\nu_e)$ &0.387&0.018&0.47&0.389&0.41&0.36\\
$\Gamma(\Lambda\to p e^-\bar\nu_e)$ &3.169&0.053&0.14&3.51&3.37&3.22\\
$\Gamma(\Sigma^-\to n e^-\bar\nu_e)$ &6.88&0.23&0.16&5.74&6.13&6.47\\
$\Gamma(\Xi^-\to \Lambda e^-\bar\nu_e)$ &3.36&0.18&0.10&2.96&2.35&2.76\\
$\Gamma(\Xi^-\to \Sigma^0e^-\bar\nu_e)$ &0.53&0.10&0.02&0.55&0.66&0.76\\
$\Gamma(\Lambda\to p \mu^-\bar\nu_\mu)$ &0.60&0.13&0.02&0.58&0.56&0.53\\
$\Gamma(\Sigma^-\to n \mu^-\bar\nu_\mu)$ &3.04&0.27&0.07&2.54&2.77&2.93\\
$\Gamma(\Xi^-\to \Lambda \mu^-\bar\nu_\mu)$ &2.1&2.1&0.03&0.80&0.65&0.76\\
\end{tabular}
\label{table:2}
\end{table}

\narrowtext
\begin{table}
\caption{Form factors $f_1$ and $g_1$ for the various semileptonic weak
beta decays. Parameter sets~3 and 4 of Table~\protect\ref{table:1}
are used.}
\begin{tabular}{ldddd}
&\multicolumn{2}{c}{Set 3}&\multicolumn{2}{c}{Set 4}\\
Decay & $f_1$ & $g_1$ & $f_1$ & $g_1$  \\
\tableline
$n\to p$            & 1.00 & 1.25 & 1.00 & 1.26 \\
$\Sigma^\pm\to \Lambda$ &--0.04 & 0.62 &--0.05 & 0.58 \\
$\Lambda\to p$      & --1.04 & --0.79 & --0.95 & --0.71 \\
$\Sigma^-\to n$       &--0.87 & 0.22 &--0.83 & 0.21 \\
$\Xi^-\to \Sigma^0$ & 0.71 & 0.86 & 0.72 & 0.86 \\
$\Xi^-\to \Lambda$    & 0.91 & 0.25 & 0.92 & 0.24 \\
\end{tabular}
\label{table:3}
\end{table}

\end{document}